# Low magnetic damping of ferrimagnetic GdFeCo alloys

Duck-Ho Kim[1†*], Takaya Okuno[1†], Se Kwon Kim[2], Se-Hyeok Oh[3], Tomoe Nishimura[1], Yuushou Hirata[1], Yasuhiro Futakawa[4], Hiroki Yoshikawa[4], Arata Tsukamoto[4], Yaroslav Tserkovnyak[2], Yoichi Shiota[1], Takahiro Moriyama[1], Kab-Jin Kim[5], Kyung-Jin Lee[3,6,7], and Teruo Ono[1,8*]

[1]Institute for Chemical Research, Kyoto University, Uji, Kyoto 611-0011, Japan

[2]Department of Physics and Astronomy, University of California, Los Angeles, California 90095, USA

[3]Department of Nano-Semiconductor and Engineering, Korea University, Seoul 02841, Republic of Korea

[4]College of Science and Technology, Nihon University, Funabashi, Chiba 274-8501, Japan

[5]Department of Physics, Korea Advanced Institute of Science and Technology, Daejeon 34141, Republic of Korea

[6]Department of Materials Science & Engineering, Korea University, Seoul 02841, Republic of Korea

[7]KU-KIST Graduate School of Converging Science and Technology, Korea University, Seoul 02841, Republic of Korea

[8]Center for Spintronics Research Network (CSRN), Graduate School of Engineering Science, Osaka University, Osaka 560-8531, Japan

† These authors contributed equally to this work.

* E-mail: kim.duckho.23z@st.kyoto-u.ac.jp, ono@scl.kyoto-u.ac.jp

We investigate the Gilbert damping parameter $\alpha$ for rare earth (RE)–transition metal (TM) ferrimagnets over a wide temperature range. Extracted from the field-driven magnetic domain-wall mobility, $\alpha$ was as low as $7.2 \times 10^{-3}$ and was almost constant across the angular momentum compensation temperature $T_A$, starkly contrasting previous predictions that $\alpha$ should diverge at $T_A$ due to vanishing total angular momentum. Thus, magnetic damping of RE-TM ferrimagnets is not related to the total angular momentum but is dominated by electron scattering at the Fermi level where the TM has a dominant damping role.

Magnetic damping, commonly described by the Gilbert damping parameter, represents the magnetization relaxation phenomenon, describing how quickly magnetization spins reach equilibrium [1–3]. Understanding the fundamental origin of the damping as well as searching for low damping materials has been a central theme of magnetism research. Several theoretical models for magnetic damping have been proposed [4–11] and compared with experiments [12–20]. Ultra-low damping was predicted in ferromagnetic alloys using a linear response damping model [11] and was demonstrated experimentally for CoFe alloys [20]. However, the majority of these studies have focused only on ferromagnetic systems.

Antiferromagnets, which have alternating orientations of their neighboring magnetic moments, have recently received considerable attention because of their potential importance for spintronic applications [21–30]. Antiferromagnetic spin systems can have much faster spin dynamics than their ferromagnetic counterparts, which is advantageous in spintronic applications [21, 25, 31–39]. However, the manipulation and control of antiferromagnets is challenging because the net magnetic moment is effectively zero. Recently, antiferromagnetic spin dynamics have been successfully demonstrated using the magnetic domain-wall (DW) dynamics in ferrimagnets with finite magnetization in the vicinity of the angular momentum compensation temperature, at which the net angular momentum vanishes [38]. This field-driven antiferromagnetic spin dynamics is possible because the time evolution of the magnetization is governed by the commutation relation of the angular momentum rather than the commutation relation of the magnetic moment.

Motivated by the aforementioned result, in this letter, we investigate the magnetic damping of ferrimagnets across the angular momentum compensation temperature, which will allow us to understand magnetic damping in antiferromagnetically coupled system. We selected rare earth (RE)–transition metal (TM) ferrimagnets for the material platforms because they have an angular momentum compensation temperature $T_A$ where antiferromagnetic spin dynamics are achieved [38, 40, 41]. The magnetic-field-driven DW motion was explored over a wide range of temperatures including $T_A$, and the Gilbert damping parameter was extracted from the measured DW mobility at each temperature by employing the collective coordinate model initially developed for ferrimagnetic spin dynamics [38]. Contrary to the previous prediction that the Gilbert damping parameter would diverge at $T_A$ due to the vanishing of the total angular momentum [42, 43], we found that the Gilbert damping parameter remained nearly constant over a wide range of temperatures across $T_A$ with the estimated value as low as $7.2 \times 10^{-3}$, which was similar to the reported values of TM-only ferromagnets [20]. These results suggested that Gilbert damping was mainly governed by electron scattering at the Fermi level, and hence, the 4f electron of the RE element, which lies far below the Fermi level, did not play an important role in the magnetic damping of RE–TM ferrimagnets.

For this study, we prepared perpendicularly magnetized ferrimagnetic GdFeCo films in which the Gd and FeCo moments were coupled antiferromagnetically. Specifically, the films were 5-nm SiN/30-nm $Gd_{23.5}Fe_{66.9}Co_{9.6}$/100-nm SiN on an intrinsic Si substrate. The GdFeCo films were then patterned into 5-μm-wide and 500-μm-long microwires with a Hall cross structure using electron beam lithography and Ar ion milling. For current injection, 100-nm Au/5-nm Ti electrodes were stacked on the wire. A Hall bar was designed to detect the DW velocity via the anomalous Hall effect (AHE).

We measured the magnetic DW motion using a real-time DW detection technique [38, 40, 41, 44, 45] [see Fig. 1(a) for a schematic]. We first applied a magnetic field of –200 mT to saturate the magnetization along the –z direction. Subsequently, a constant perpendicular magnetic field $\mu_0 H$, which was lower than the coercive field, was applied along +z direction. Next, a d.c. current was applied along the wire to measure the anomalous Hall voltage. Then, a current pulse (12 V, 100 ns) was injected through the writing line to nucleate the DW in the wire. The created DW was moved along the wire and passed through the Hall bar because of the presence of $\mu_0 H$. The DW arrival time was detected by monitoring the change in the Hall voltage using a real-time oscilloscope. The DW velocity could then be calculated from the arrival time and the travel distance between the writing line and Hall bar (500 μm).

Figure 1(b) shows the averaged DW velocity $\langle v \rangle$ as a function of the perpendicular magnetic field $\mu_0 H$ for several temperatures $T^*$. Here, we used the d.c. current density of $|J| = 1.3 \times 10^{10}$ A/m² to measure the AHE change due to DW motion. Note that $T^*$ is an elevated temperature that considers Joule heating by d.c. current [46]. To eliminate the undesired current-induced spin-transfer-torque effect, we averaged the DW velocity for $+J$ and $-J$, i.e., $\langle v \rangle = [v(+J) + v(-J)]/2$. Figure 1(b) shows that $\langle v \rangle$ increases linearly with $\mu_0 H$ for all $T^*$. Such linear behavior can be described by $\langle v \rangle = \mu[\mu_0 H - \mu_0 H_0]$, where $\mu$ is the DW mobility and $\mu_0 H_0$ is the correction field, which generally arises from imperfections in the sample or complexities of the internal DW structure [47, 48]. We note that $\mu_0 H_0$ can also depend on the temperature dependence of the magnetic properties of ferrimagnets [45]. Figure 1(c) shows $\mu$ as a function of $T^*$ at several current densities ($|J| = 1.3$, 1.7, and 2.0 $\times 10^{10}$ A/m²). A sharp peak clearly occurs for $\mu$ at $T^* = 241.5$ K irrespective of $|J|$. The drastic increase of $\mu$ is evidence of antiferromagnetic spin dynamics at $T_A$, as demonstrated in our previous report [38, 40, 41].

The obtained DW mobility was theoretically analyzed as follows. The DW velocity of ferrimagnets in the precessional regime is given by [38, 39]

$$V = \lambda \alpha \frac{(s_1 + s_2)(M_1 - M_2)}{[\alpha(s_1 + s_2)]^2 + (s_1 - s_2)^2} \mu_0 H, \tag{1}$$

where $V$ is the DW velocity, $\lambda$ is the DW width, $\mu_0 H$ is the perpendicular magnetic field, $\alpha$ is the Gilbert damping parameter, $M_i$ and $s_i$ are the magnetization and the spin angular momentum of one sublattice, respectively. The spin angular momentum densities are given by $s_i = M_i/\gamma_i$ [49], where $\gamma_i = g_i \mu_B/\hbar$ is the gyromagnetic ratio of lattice $i$, $g_i$ is the Landé g factor of lattice $i$, $\mu_B$ is the Bohr magneton, and $\hbar$ is the reduced Plank's constant. The Gilbert damping is in principle different for two sublattices, but for simplicity, we assume that it is the same, which can be considered as the average value of the damping parameters for the two sublattices weighted by the spin angular momentum density. We note that this assumption does not alter our main conclusion: low damping and its insensitivity to the temperature. Equation (1) gives the DW mobility $\mu$ as $\lambda \alpha (s_1 + s_2)(M_1 - M_2)/\{[\alpha(s_1 + s_2)]^2 + (s_1 - s_2)^2\}$, which can be rearranged as

$$\mu(s_1 + s_2)^2 \alpha^2 - \lambda(s_1 + s_2)(M_1 - M_2)\alpha + \mu(s_1 - s_2)^2 = 0 \tag{2}$$

Using Eq. (2) to find the solution of $\alpha$, we find

$$\alpha_\pm = \frac{\lambda(M_1 - M_2) \pm \sqrt{[\lambda^2(M_1 - M_2)^2 - 4\mu^2(s_1 - s_2)^2]}}{2\mu(s_1 + s_2)}. \tag{3}$$

Equation (3) allows us to estimate $\alpha$ for the given $\mu$. We note that for each value of $\mu$, $\alpha$ can have two values, $\alpha_+$ and $\alpha_-$ because of the quadratic nature of Eq. (2). Only one of these two solutions is physically sound, which can be obtained using the following energy dissipation analysis.

The energy dissipation (per unit cross section) through the DW dynamics is given by $P = 2\alpha(s_1 + s_2)V^2/\lambda + 2\alpha(s_1 + s_2)\lambda\Omega^2$ [38, 39], where $\Omega$ is the angular velocity of the DW. The first and the second terms represent the energy dissipation through the translational and angular motion of the DW, respectively. In the precessional regime, the angular velocity is proportional to the translational velocity: $\Omega = (s_1 - s_2)V/\alpha(s_1 + s_2)\lambda$. Replacing $\Omega$ by the previous expression yields $P = \eta V^2$ where $\eta = 2(M_1 - M_2)/\mu$ is the viscous coefficient for the DW motion:

$$\eta = \frac{2}{\lambda}\left\{\alpha(s_1 + s_2) + \frac{(s_1 - s_2)^2}{\alpha(s_1 + s_2)}\right\}. \tag{4}$$

The first and the second terms in parenthesis capture the contributions to the energy dissipation from the translational and angular dynamics of the DW, respectively. The two solutions for the Gilbert damping parameter, $\alpha_+$ and $\alpha_-$, can yield the same viscous coefficient $\eta$. The case of the equal solutions, $\alpha_+ = \alpha_-$, corresponds to the situation when the two contributions are identical: $\alpha_\pm = (s_1 - s_2)/(s_1 + s_2)$. For the larger solution $\alpha = \alpha_+$, the energy dissipation is dominated by the first term, i.e., through the translational DW motion, which should be the case in the vicinity of $T_A$ where the net spin density $(s_1 - s_2)$ is small and thus the angular velocity is negligible. For example, at exact $T_A$, the larger solution $\alpha_+$ is the only possible solution because the smaller solution is zero, $\alpha_- = 0$, and thus unphysical. For the smaller solution $\alpha = \alpha_-$, the dissipation is dominated by the second term, i.e., through the precessional motion, which should describe cases away from $T_A$. Therefore, in the subsequent analysis, we chose the larger solution $\alpha_+$ in the vicinity of $T_A$ and the smaller solution $\alpha_-$ far away from $T_A$ and connected the solution continuously in between.

The other material parameters such as $M_1$, $M_2$, $s_1$, and $s_2$ are estimated by measuring the net magnetic moment of GdFeCo film, $|M_{\text{net}}|$, for various temperatures. Because $M_{\text{net}}$ includes contributions from both the Gd and FeCo sub-moments, the sub-magnetic moments, $M_1$ and $M_2$, could be decoupled based on the power law criticality [see details in refs. 38, 40]. The spin angular momentums, $s_1$ and $s_2$, were calculated using the known Landé g factor of FeCo and Gd (the Landé g factor of FeCo is 2.2 and that of Gd is 2.0) [50–52].

Figures 2(a)–(c) show the temperature-dependent DW mobility $\mu$, sub-magnetic moment $M_i$, and sub-angular momentum $s_i$, respectively. Here, we used the relative temperature defined as $\Delta T = T^* - T_A$ to investigate the Gilbert damping near $T_A$. The Gilbert damping parameter $\alpha$ was obtained based on Eq. (3) and the information in Fig. 2(a)–(c). Figure 2(d) shows the resulting values of $\alpha_\pm$ as a function of $\Delta T$. For $\Delta T_1 < \Delta T < \Delta T_2$, $\alpha_+$ is nearly constant, while $\alpha_-$ varies significantly. For $\Delta T < \Delta T_1$ and $\Delta T > \Delta T_2$, on the other hand, $\alpha_-$ is almost constant, while $\alpha_+$ varies significantly. At $\Delta T = \Delta T_1$ and $\Delta T = \Delta T_2$, the two solutions are equal, corresponding to the aforementioned case when the energy dissipation through the translational and angular motion of the DW are identical.

The proper damping solution can be selected by following the guideline obtained from the above analysis. For $\Delta T_1 < \Delta T < \Delta T_2$, which includes $T_A$, the energy dissipation should be dominated by the translational motion, and thus $\alpha_+$ is a physical solution. Note also that $\alpha_-$ becomes zero at $T_A$, which results in infinite DW mobility in contradiction with the experimental observation. For $\Delta T < \Delta T_1$ and $\Delta T > \Delta T_2$, where the energy dissipation is dominated by the angular motion of the DW, $\alpha_-$ is the physical solution.

Figure 3 shows the resultant Gilbert damping parameter in all tested temperature ranges. The Gilbert damping parameter was almost constant across $T_A$ with $\alpha = 7.2 \times 10^{-3}$ (see the dotted line in Fig. 3). This result is in stark contrast to the previous prediction. In ref. [42], Stanciu *et al*. investigated the temperature dependence of the effective Gilbert damping parameter based on a ferromagnet-based model and found that the damping diverged at $T_A$. Because they analyzed the magnetic resonance in ferrimagnetic materials based on a ferromagnet-based model, which cannot describe the antiferromagnetic dynamics at $T_A$ at which the angular momentum vanishes, it exhibits unphysical results. However, our theoretical analysis for field-driven ferromagnetic DW motion based on the collective coordinate approach can properly describe both the antiferromagnetic dynamics in the vicinity of $T_A$ and the ferromagnetic dynamics away from $T_A$ [38]. Therefore, the unphysical divergence of the Gilbert damping parameter at $T_A$ is absent in our analysis.

Our results, namely the insensitivity of damping to the compensation condition and its low value, have important implications not only for fundamental physics but also for technological applications. From the viewpoint of fundamental physics, nearly constant damping across $T_A$ indicates that the damping is almost independent of the total angular momentum and is mostly determined by electron spin scattering near the Fermi level. Specifically, our results suggest that the 4f electrons of RE elements, which lie in a band far below the Fermi level, do not play an important role in the magnetic damping of RE-TM ferrimagnets, whereas the 3d and 4s bands of TM elements have a governing role in magnetic damping. This result is consistent with the recently reported theoretical and experimental results in FeCo alloys [20]. From the viewpoint of practical application, we note that the estimated damping of $\alpha = 7.2 \times 10^{-3}$ is the upper limit, as the damping estimated from DW dynamics is usually overestimated due to disorders [53]. The obtained value of the Gilbert damping parameter is consistent with our preliminary ferromagnetic resonance (FMR) measurements. The experimental results from FMR measurements and the corresponding theoretical analysis will be published elsewhere. This low value of the Gilbert damping parameter suggests that ferrimagnets can serve as versatile platforms for low-dissipation high-speed magnetic devices such as spin-transfer-torque magnetic random-access memory and terahertz magnetic oscillators.

In conclusion, we investigated the field-driven magnetic DW motion in ferrimagnetic GdFeCo alloys over a wide range of temperatures across $T_A$ and extracted the Gilbert damping parameter from the DW mobility. The estimated Gilbert damping parameter was as low as $7.2 \times 10^{-3}$ and almost constant over the temperature range including $T_A$, which is in stark contrast to the previous prediction in that the Gilbert damping parameter would diverge at $T_A$ due to the vanishing total angular momentum. Our finding suggests that the magnetic damping of RE-TM ferrimagnets is not related to the total angular momentum but is mostly governed by the scattering of electrons at the Fermi level where the TM element has a dominant role for the magnetic damping.

**Figure Captions**

Figure 1(a) Schematic illustration of the GdFeCo microwire device. (b) The averaged DW velocity $\langle v \rangle$ as a function of the perpendicular magnetic field $\mu_0 H$ for several temperatures $T^*$ (202, 222, 242, 262, and 282 K). The dots indicate the best linear fits. (c) The DW mobility $\mu$ as a function of $T^*$ at several current densities ($|J|$ =1.3, 1.7, and 2.0 ×10$^{10}$ A/m²).

Figure 2 The temperature-dependent (a) DW mobility $\mu$, (b) sub-magnetic moment $M_i$, and (c) sub-angular momentum $s_i$. Here, we use the relative temperature defined as $\Delta T = T^* - T_A$. (d) The Gilbert damping parameter $\alpha_\pm$ as a function of $\Delta T$. Here, we use $\lambda$ =15 nm for proper solutions of Eq. (3).

Figure 3 The resultant Gilbert damping parameter $\alpha$ in all tested temperature ranges.

**Acknowledgements**

This work was supported by the JSPS KAKENHI (Grant Numbers 15H05702, 26103002, and 26103004), Collaborative Research Program of the Institute for Chemical Research, Kyoto University, and R & D project for ICT Key Technology of MEXT from the Japan Society for the Promotion of Science (JSPS). This work was partly supported by The Cooperative Research Project Program of the Research Institute of Electrical Communication, Tohoku University. D.H.K. was supported as an Overseas Researcher under the Postdoctoral Fellowship of JSPS (Grant Number P16314). S.H.O. and K.J.L. were supported by the National Research Foundation of Korea (NRF-2015M3D1A1070465, 2017R1A2B2006119) and the KIST Institutional Program (Project No. 2V05750). S.K.K. was supported by the Army Research Office under Contract No. W911NF-14-1-0016. K.J.K. was supported by the National Research Foundation of Korea (NRF) grant funded by the Korea Government (MSIP) (No. 2017R1C1B2009686).

**Competing financial interests**

The authors declare no competing financial interests.

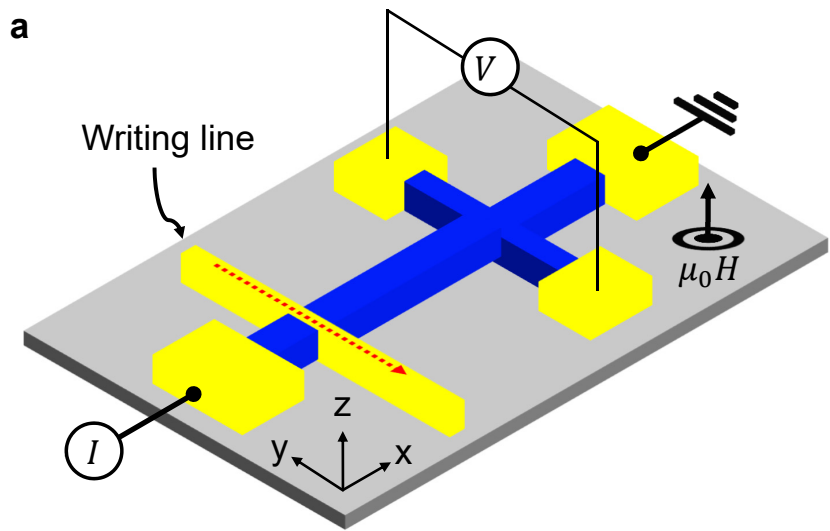
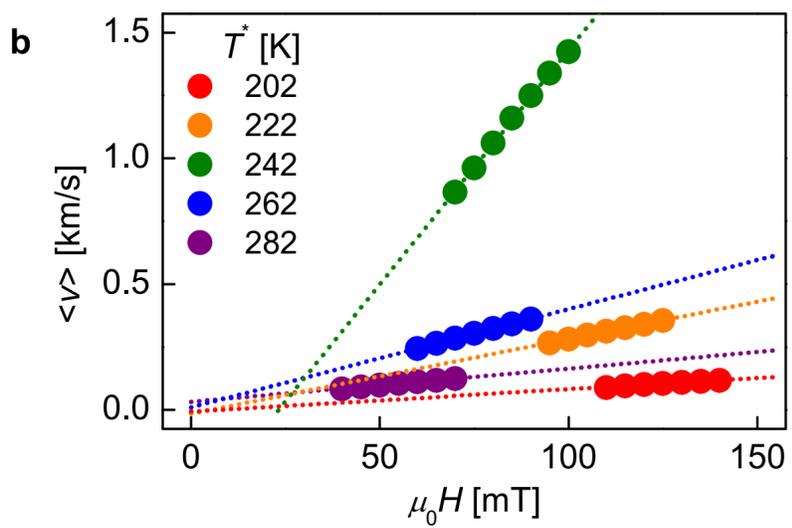
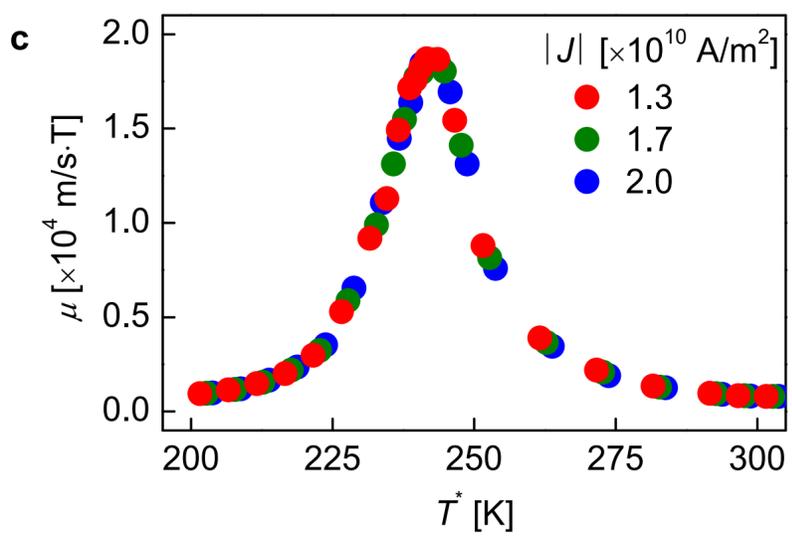

**Figure 1**

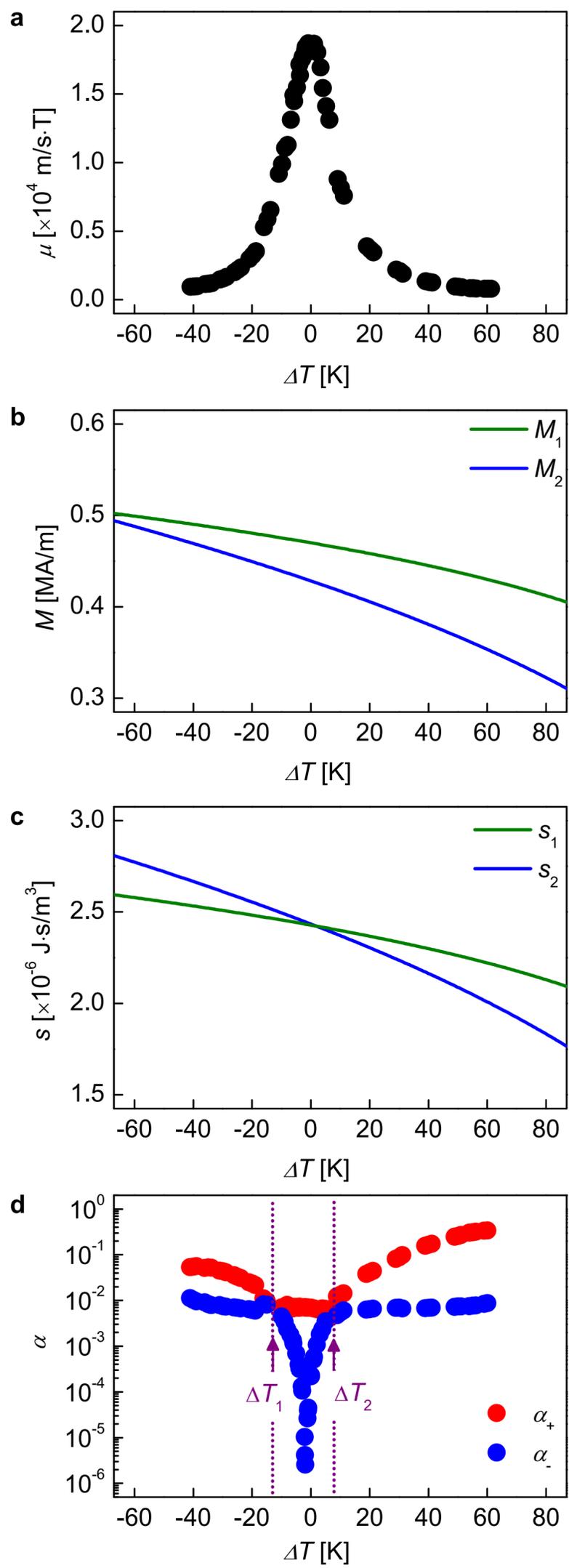

Figure 2

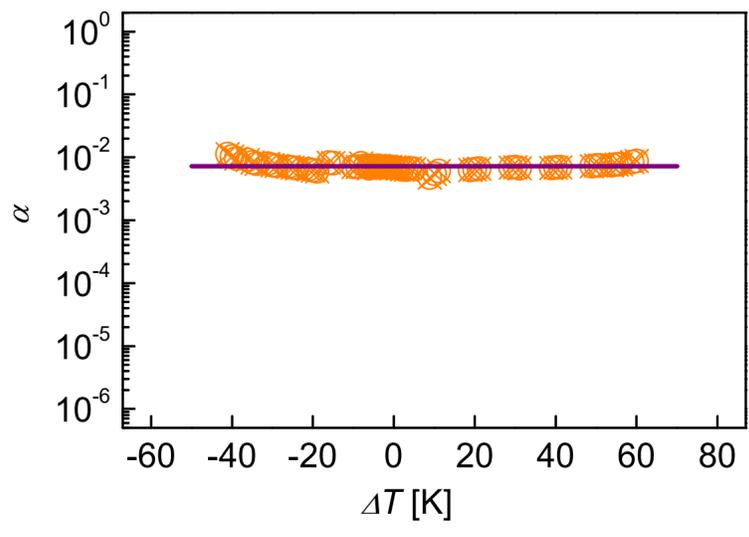

Figure 3